\documentclass[
    ,final            
  ]
  {aipproc}

\layoutstyle{6x9}

\begin{document}

\title{Joint X-ray/Sunyaev-Zel'dovich Analysis of the Intra-Cluster Medium}


\author{Kaustuv Basu}{
  address={Max-Planck-Institut f\"ur Radioastronomie, Bonn, Germany}
  ,altaddress={Argelander Institut f\"ur Astronomie, Bonn, Germany}
}

\author{Martin W. Sommer}{
  address={Argelander Institut f\"ur Astronomie, Bonn, Germany}
}

\author{Yu-Ying Zhang}{
  address={Argelander Institut f\"ur Astronomie, Bonn, Germany}
}

\begin{abstract}

We present results from a joint X-ray/Sunyaev-Zel'dovich modeling of the intra-cluster gas using XMM-Newton and APEX-SZ imaging data. The goal is to study the physical properties of the intra-cluster gas with a non-parametric de-projection method that is, aside from the assumption of spherical symmetry, free from modeling bias. We demonstrate a decrease of gas temperature in the cluster outskirts, and also measure the gas entropy profile, both of which are obtained for the first time independently of X-ray spectroscopy, using Sunyaev-Zel'dovich and X-ray imaging data. The contribution of the APEX-SZ systematic uncertainties in measuring the gas temperature at large radii is shown to be small compared to the XMM-Newton and Chandra systematic spectroscopic errors.

\end{abstract}

\maketitle


\section{Introduction}

Accurately determining the thermodynamic state of the intra-cluster medium (ICM) out to large radii in galaxy clusters is critical for understanding the link between the total cluster mass and X-ray observables. 
For over a decade, observations of the thermal Sunyaev-Zel'dovich Effect (tSZE, hereafter simply SZ; Sunyaev \& Zel'dovich 1970) have been considered as a promising complement to X-ray observations for modeling the ICM in galaxy clusters, yet only recently has it been possible to make meaningful de-projections of gas temperature and density profiles using  SZE imaging data from multi-pixel bolometer arrays. The APEX-SZ experiment (Dobbs et al. 2006, Halverson et al. 2009) employs one of the first such powerful multi-pixel Transition-Edge Sensor (TES) bolometer cameras, enabling a joint analyses of the ICM properties using SZE and X-ray data. The first results of such analysis have been published by Nord et al. (2009) for the cluster Abell 2163, and Basu et al. (2010) for the prototypical relaxed cluster Abell 2204.

\section{Methods}

\begin{figure}[t]
\hspace*{-8mm}
\includegraphics[width=9cm, trim=30 10 30 30,clip]{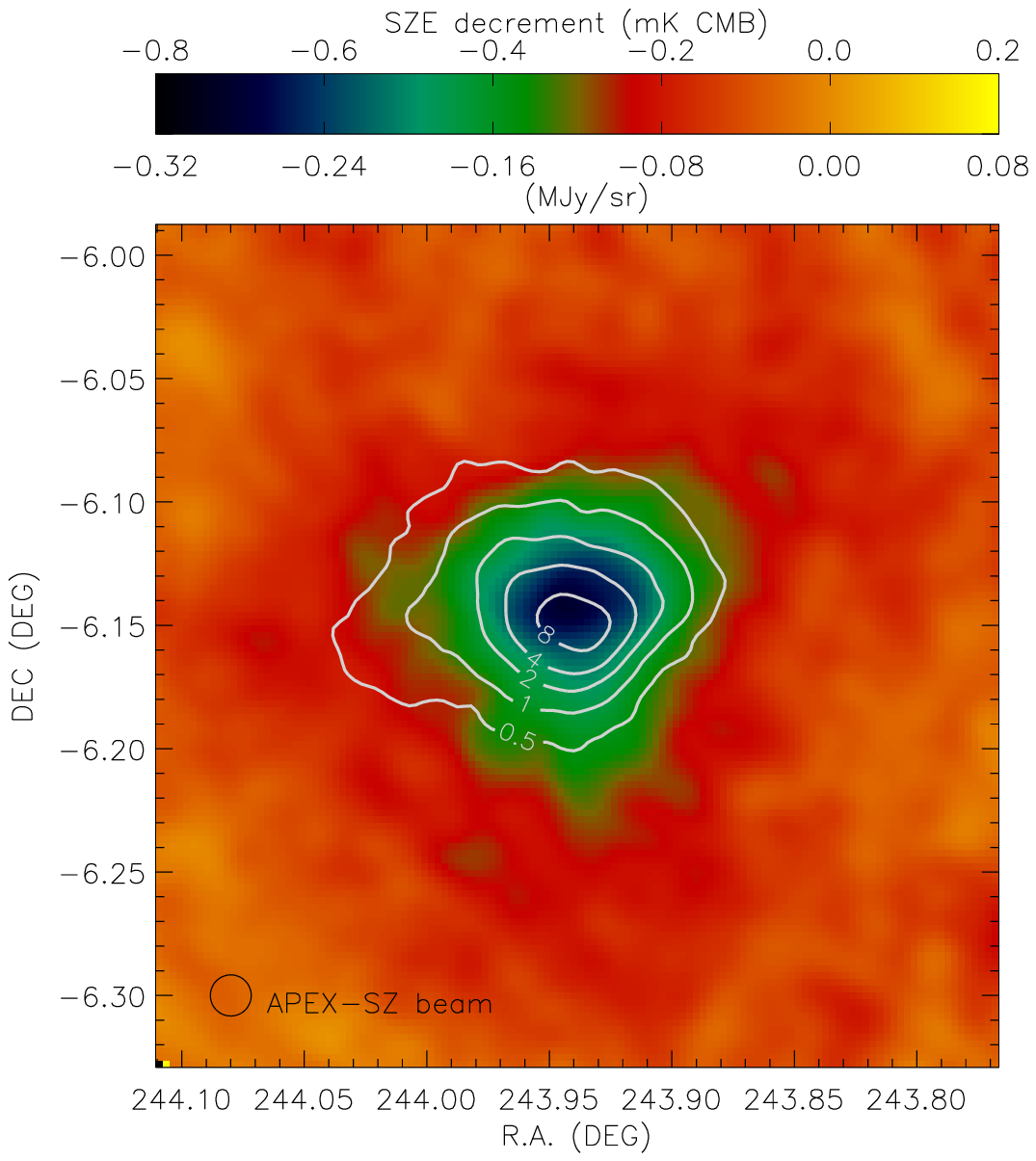}%
\includegraphics[width=5.5cm, trim=0 0 0 0,clip]{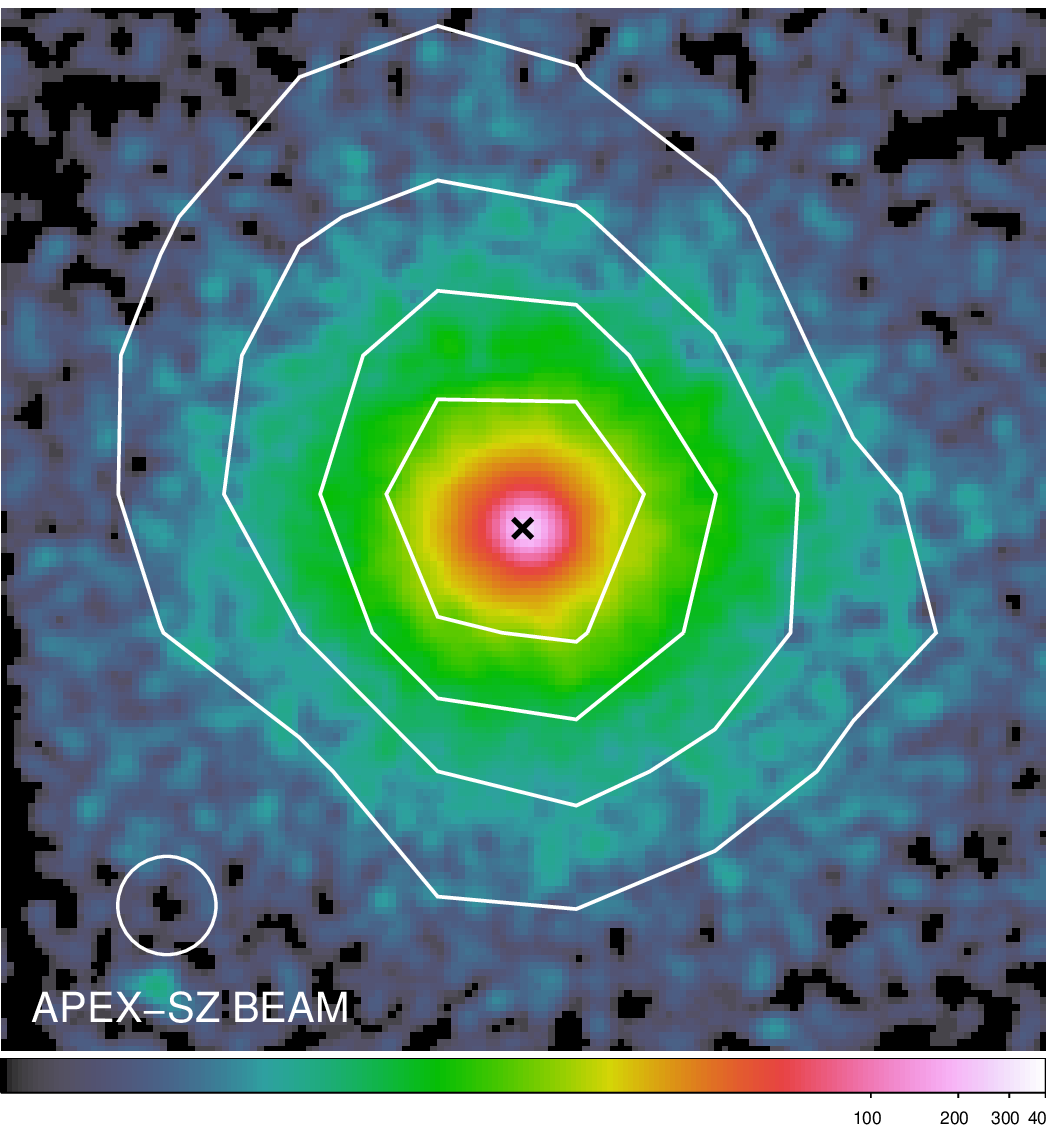}
\caption{\textit{Left panel:} APEX-SZ map of Abell 2163 at 150 GHz, overlaid with XMM-Newton X-ray brightness contours in the 0.5-2~keV band. \textit{Right panel:}
XMM-Newton MOS1 image of Abell 2204 overlaid with the contours from its APEX-SZ measurement. The SZ contour steps are -2, -4, -6 and -8 $\sigma$ with a spatial resolution of one arcminute and map size of $10\times 10$~arcmin. 
The black cross marks the flux weighted X-ray center, which matches with the peak location of the SZE map within the pointing accuracy of the APEX telescope.}
\label{fig:xmap}
\end{figure}

We use publicly available X-ray imaging data in the 0.7--2 keV energy band from the \textit{XMM-Newton} observatory, and our Sunyaev-Zel'dovich effect imaging data at 150 GHz from the APEX-SZ experiment, to de-project the density and temperature profiles in clusters. The details of the X-ray and SZ data reductions are described in Nord et al. (2009) and Basu et al. (2010). In Fig.\ref{fig:xmap} we show examples of the SZ and X-ray imaging data used in our analysis, for the clusters Abell 2163 and Abell 2204. 

The three-dimensional (de-projected) density and temperature profiles are obtained directly using Abel's integral inversion method, originally proposed by Silk \& White (1978) for joint X-ray/SZ analyses of galaxy clusters. 
The high sensitivity and one arcminute resolution of the APEX-SZ images makes it possible to apply this method to perform a non-parametric analysis of the ICM properties in a real cluster. 
The uncertainties in the radial temperature profile are as yet dominated by the statistical uncertainties in the SZ measurement.

\section{Results}

\begin{figure}[t]
\centering
\includegraphics[width=8cm]{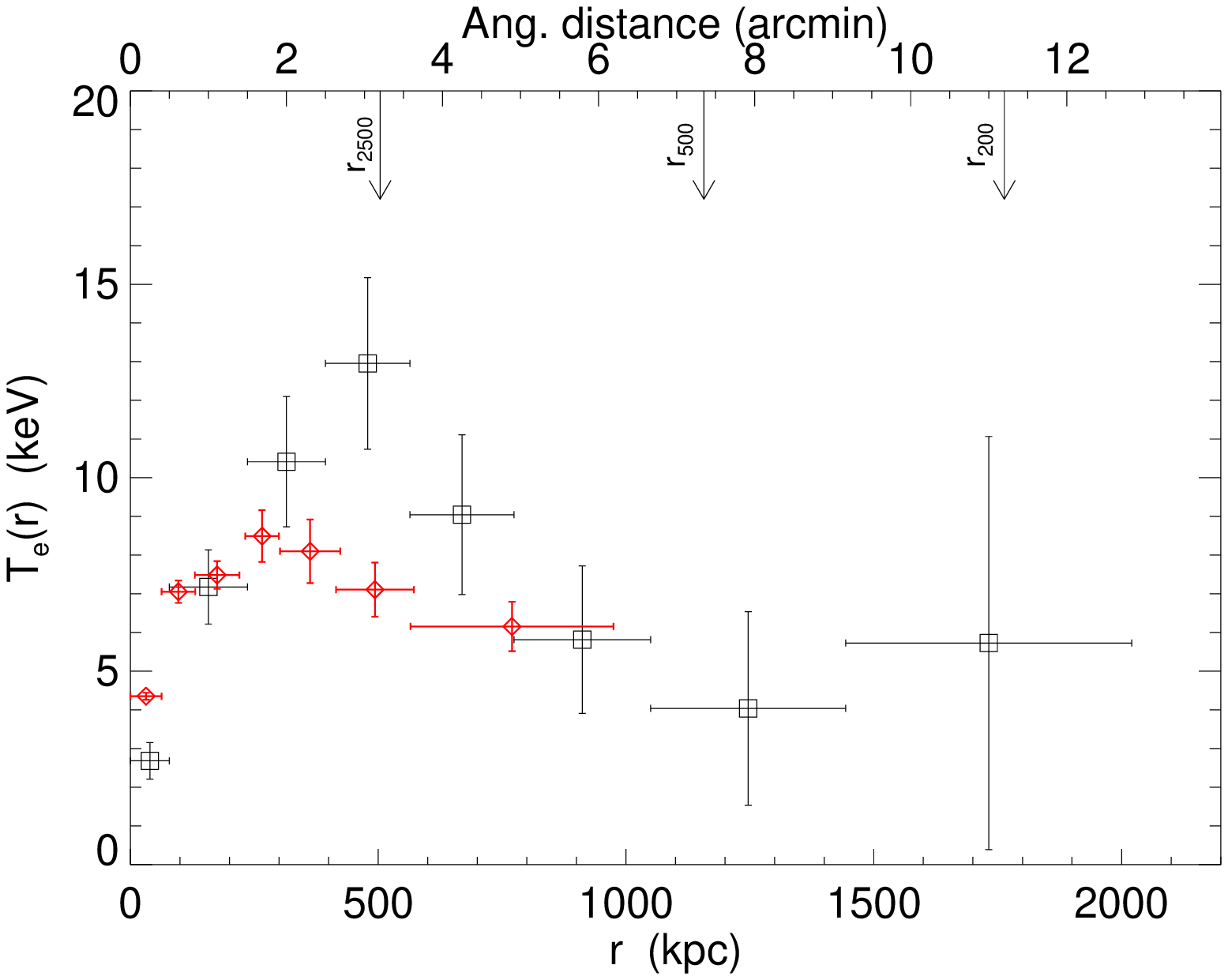}%
\includegraphics[width=8cm]{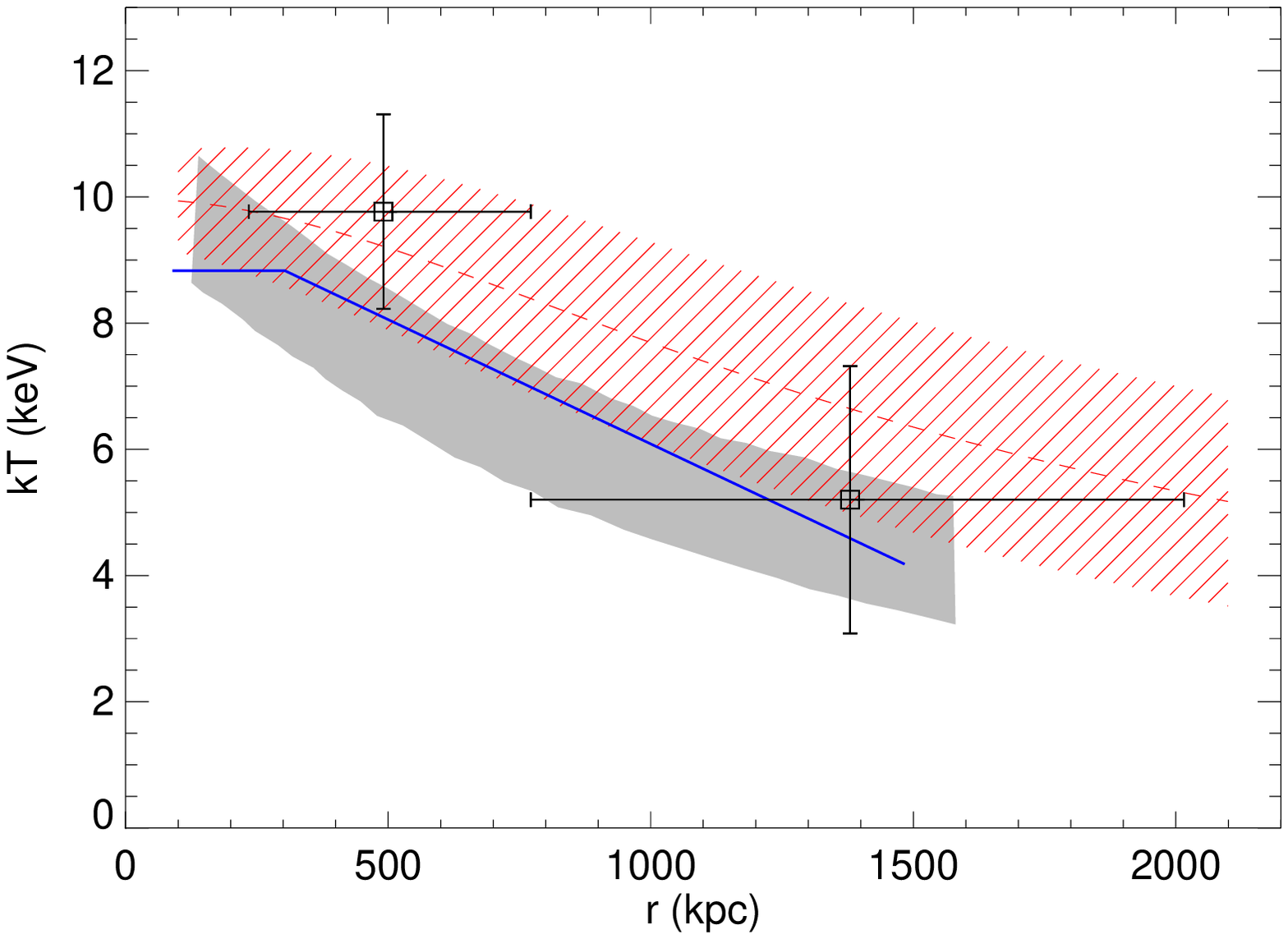}
\caption{ \textit{Left panel:} De-projected temperature values and their $1\sigma$ statistical errors from the SZE measurement of the relaxed cluster Abell 2204. Over-plotted data points (red diamonds) are from the  \textit{XMM-Newton} analysis by Zhang et al. (2008). \textit{Right panel:} SZ measurement of the temperature drop in the outskirts of Abell 2204, shown in comparison with the results from cluster simulations by Hallman et al. 2007 (red dashed region),  measurements of cooling core clusters from ASCA (gray shaded area) and \textit{Chandra} (blue solid line).}
\label{fig:denstemp}
\end{figure}

The primary outputs of the de-projection analysis are the three-dimensional density and temperature profiles, from which the radial profiles of gas entropy and gas mass are obtained. The gas density profiles of the modeled clusters agree well with the X-ray derived isothermal $\beta$-models (Cavaliere \& Fusco-Femiano 1978), confirming that in the 0.5--2 keV energy band the X-ray surface brightness is practically independent of the gas temperature. The gas temperature profile within $r_{500}$ in Abell 2163 is consistent with being isothermal, whereas for the cool core cluster Abell 2204 a clear drop in the gas temperature near the cluster center is seen. By re-binning the data for Abell 2204, we can also confirm a drop in the gas temperature in the cluster outskirts at 98\% confidence level, which is obtained for the first time without using X-ray spectroscopy. The results for the  temperature modeling in Abell 2204 are shown in Fig.\ref{fig:denstemp}.

\begin{figure}[t]
\centering
\includegraphics[width=8cm]{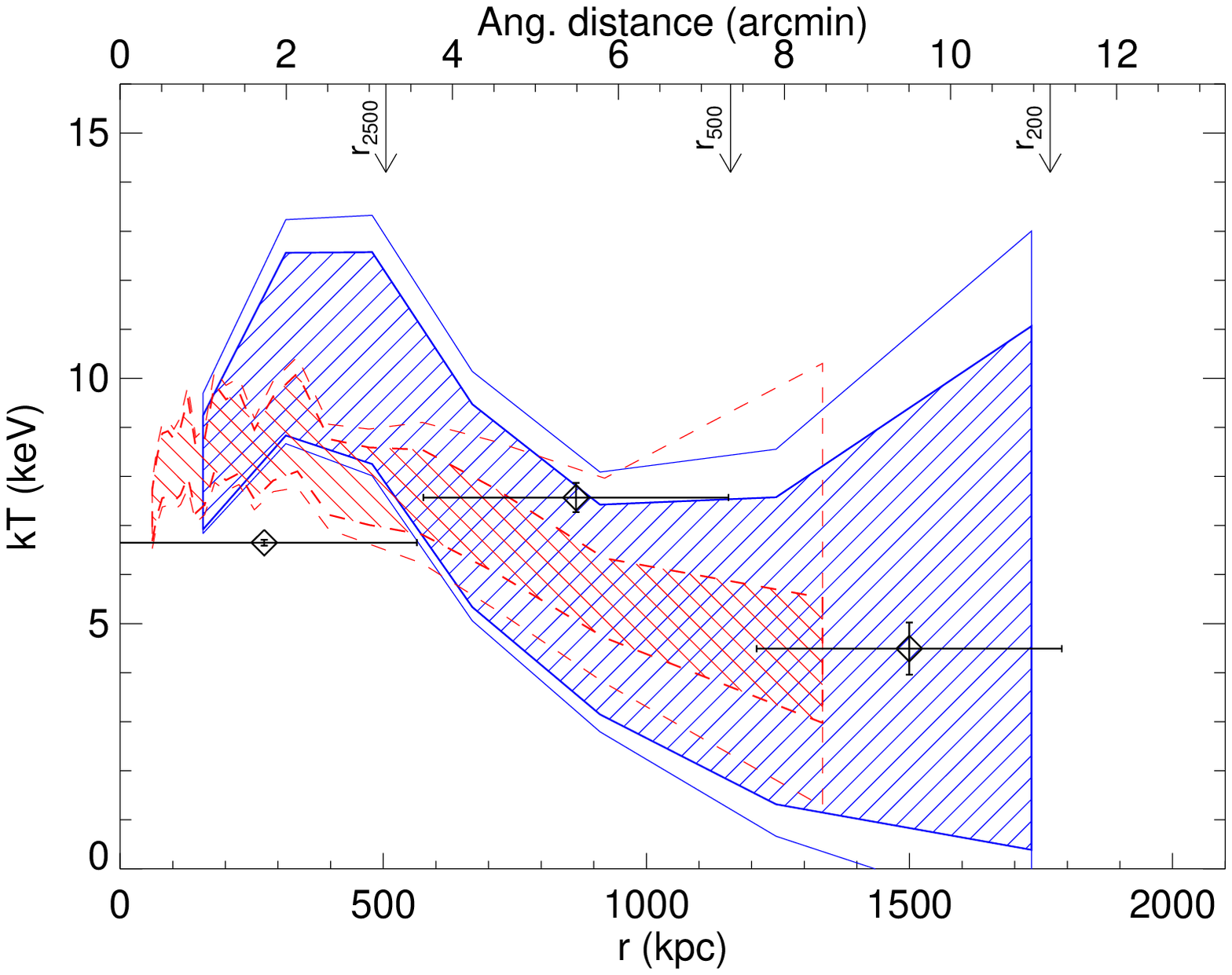}%
\includegraphics[width=8cm]{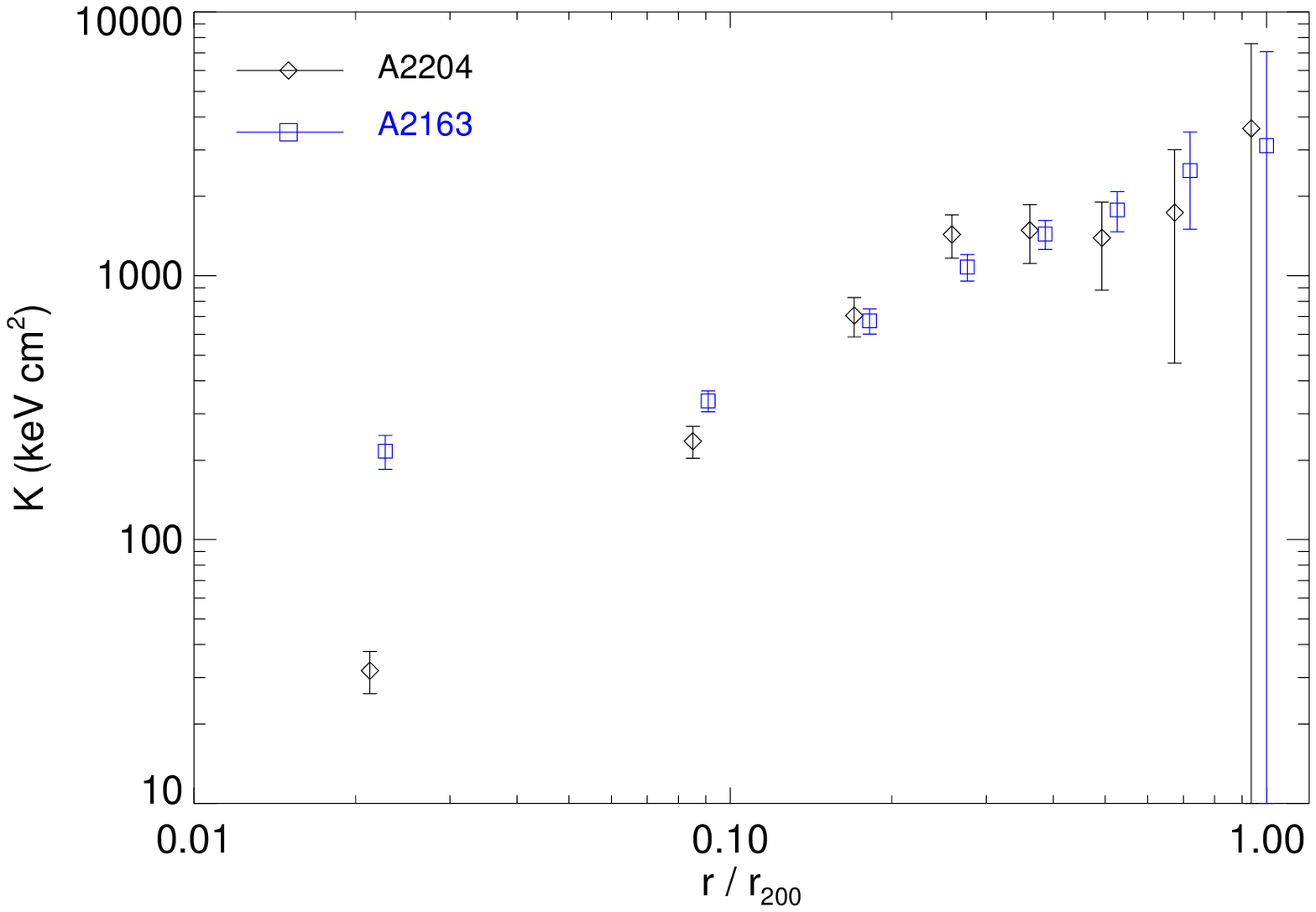}
\caption{\textit{Left panel:} Comparison of the projected gas temperature profile in Abell 2204 deduced from APEX-SZ data (blue, solid boundaries) with X-ray spectral measurements from \textit{Chandra} data (red, dashed boundaries). The hatched regions show the $1\sigma$ statistical uncertainties in each measurement, on top of which the total uncertainties in each method are overplotted combining statistical and systematic errors in quadrature. The three data points (black diamonds) are the \textit{Suzaku} spectroscopic measurements by Reiprich et al. (2009). \textit{Right panel:} Comparison between the entropy profiles for Abell 2204 (black diamonds) and A2163 (blue squares). The higher entropy value in the central region of A2163 indicates the non-relaxed state of this cluster.}
\label{fig:randsys}
\end{figure}

To compare the SZ-derived temperature profile in Abell 2204 with that obtained from X-ray spectroscopic measurements, we obtained the ``spectroscopic-like'' temperature following Mazzotta et al. (2004). We also re-analyzed the \textit{Chandra} data (total exposure time 88 ks) using the latest calibration updates (CALDB 4). The result for this comparison is shown in Fig.\ref{fig:randsys}, together with the \textit{Suzaku} measurements for this same cluster.  The \textit{Suzaku} measurements in the inner bins are affected by its large PSF (Reiprich et al. 2009). At $r_{500}$ the \textit{Chandra} measurements are dominated by systematic uncertainties due to the background modeling; beyond that radius it is impossible to put meaningful constraints  using the current \textit{Chandra} data. In comparison, the uncertainties on the SZE-derived temperatures are dominated by the statistical errors, out to $r_{200}$. The very low and stable particle background in the \textit{Suzaku} orbit makes its spectroscopic temperature measurements at large radii far superior. However, its large PSF is a problem for modeling clusters at higher redshifts $(z>0.2)$.

We can compute the gas entropy profile, defined as $K= T_e\ n_e^{-2/3}$, directly from the de-projected density and temperature profiles. A comparison between the entropy values in Abell 2163 and Abell 2204 shows a clear difference within the central 200 kpc, with the core entropy in Abell 2163 being roughly an order of magnitude higher (Fig.\ref{fig:randsys}). Outside the core the entropy profile agrees with the power law $K(r) \propto r^{1.1}$ expected from self-similar cluster models. The higher value of the ``central entropy floor'' in Abell 2163 most likely indicates its merging nature. This entropy difference is again seen for the first time from SZ-derived temperature measurements independently of X-ray spectroscopy.


\section{Conclusions}

The potential for joint X-ray/SZ analysis of the ICM properties is shown with the help of APEX-SZ data, which allows for a non-parametric modeling of the temperature and density profiles out to the cluster virial radius. Apart from the assumption of spherical symmetry, the de-projection method is free from modeling biases. The uncertainties in the gas temperature measurements near the virial radii is dominated by the statistical uncertainties in the SZ measurements. A demonstration of the decreasing gas temperature in the cluster outskirts, and also the measurement of gas entropy profiles, are made from the SZ and X-ray imaging data without the use of X-ray spectroscopy.

\begin{theacknowledgments}

The authors thank H. Eckmiller, V. Jaritz, T. H. Reiprich, and the members of the APEX-SZ collaboration, for their contributions to this work.

\end{theacknowledgments}


\end{document}